# Estimating Model Error Covariances with Artificial Neural Networks

Massimo Bonavita and Patrick Laloyaux

*ECMWF, Reading, UK*

## ABSTRACT

Methods to deal with systematic model errors are an increasingly important component of modern data assimilation systems and their effectiveness has increased in recent years thanks to advances in methodology and the quality and density of the global observing system. The weak constraint 4D-Var assimilation algorithm employed at ECMWF is well suited to the estimation and correction of model errors as they are explicitly accounted for in the cost function. This has led to significant improvements in recent years to the accuracy of stratospheric analyses. One question that remains open is about the estimation of the model error covariance matrix to use in weak constraint 4D-Var. Encouraged by the promising results we have obtained in the recent past through the use of Artificial Neural Networks (ANNs) to estimate slowly-varying model errors in the ECMWF assimilation cycle, we explore in this work the use of ANNs to sample the model error distribution and provide an alternative way to construct a model error covariance matrix. Results from the application of the new model error covariance in cycling assimilation experiments are described and implications for further developments of the ECMWF data assimilation system are discussed.

# 1) Introduction

Chaotic growth of errors in the initial conditions and model deficiencies are the main limiters of forecast skill in Numerical Weather Prediction (NWP) and, more generally, in Earth System Prediction (ESP). Data assimilation methodology and applications have traditionally been more focussed on the estimation and correction of the first type of errors (also known as predictability errors). However, with the rapid increase in the number and accuracy of observations available for Earth system initialisation (ECMWF, 2020) there is an increasing awareness that model errors are becoming a more significant component of the analysis and forecast error budgets.

Model deficiencies arise from several concomitant and interrelated factors (e.g., lack of resolution, numerical discretisation errors, missing or approximate representation of physical processes, etc.) which typically lead to systematic, state-dependent deviations of the model evolution from the truth. While initial condition uncertainties and predictability errors can be effectively characterised through ensemble data assimilation and forecasting techniques (e.g., Bonavita et al., 2012, Leutbecher et al., 2017), the treatment of systematic, time-correlated model errors has received considerably less attention, at least in NWP applications.

In a data assimilation framework, a necessary pre-condition to deal successfully with model error is to be able to access a realistic estimate of the model error covariance matrix. This problem has received attention in both the statistical and meteorological literature, and some of the more relevant results are reviewed below.

One general class of methods uses innovations (i.e., differences between observations and model equivalents) to derive estimates of the model error covariance and, usually at the same time, the observation error covariance matrix (see Tandeo et al., 2020, for a recent review; Todling 2015; Bowler, 2017), as model and observation errors are recognised as the fundamental statistical inputs needed to achieve optimal results in a data assimilation system. Different types of innovations (O-B, O-A, time-lagged innovations, etc.) are used in these methods; also, different types of statistical approaches are brought to bear, i.e., Bayesian vs frequentist, method of moments vs maximum likelihood estimation, etc. One common thread in these works is the realisation that in order to be successful all these methods need to rely on specific (and often hard to justify) assumptions about the structure and statistical properties of either or both model error and observation error covariances (e.g., stationarity, isotropy, statistical independence, etc.). A significant drawback of these methods is that they provide error covariance estimates in the space of the observations. This poses both scientific questions (e.g., different model error statistics in data rich vs data poor regions or regions observed with different observation types) and practical difficulties, mainly due to the need to transform model error covariance error estimates from observation space to state/control vector space. For these reasons the above methods have seen limited application in the estimation of model error statistics in realistic NWP systems, except for their more limited use in guiding the tuning process of background forecast error variances used in the assimilation cycle (see Tandeo et al., 2020 and references therein). Another general class of methods (first proposed by Daley, 1992; see also Boisserie et al., 2014) relies on the explicit partition of forecast errors in: a) errors from the evolution of uncertain initial conditions (i.e., predictability errors), and b) model errors, i.e., errors that would arise from model imperfections even in the presence of perfect initial conditions. This distinction is, for example, formalised in the evolution equation of the state error covariance in the Kalman Filter (Kalman, 1960):

$$\mathbf{P}_t^b = \mathbf{M}\mathbf{P}_{t-1}^a\mathbf{M}^T + \mathbf{Q}^t, \qquad (1)$$

where the state error covariance matrix of the background forecast state ($\mathbf{P}_t^b$) is written as the sum of the linearly evolved analysis error covariance from the previous analysis update ($\mathbf{M}\mathbf{P}_{t-1}^a\mathbf{M}^T$; $\mathbf{M}$ is the linear/linearised prognostic model) and a model error covariance $\mathbf{Q}_t$. In this framework the estimate of $\mathbf{Q}_t$ is derived from the explicit estimation of the analysis error covariance and its evolution, plus the estimation of the total forecast error statistics. The estimation of the analysis error covariance matrix is a particularly difficult aspect of this methodology. Authors have, for example, sampled this quantity using an Ensemble of Data Assimilations (EDA, Houtekamer et al., 1996; Fisher, 2003; Bonavita et al., 2012), but it is obvious that results crucially depend on the assumptions used to sample observation and model errors in the EDA itself and can thus cause undesirable feedback loops. On the other hand, one advantage of the method is that it provides estimates of the desired covariances in model space, which facilitates application in realistic NWP and Ocean data assimilation systems. Weak Constraint 4D-Var (WC-4DVar, Trèmolet, 2006) is an extension of 4D-Var which explicitly takes model error into account in the solution of the 4D-Var minimisation problem. This makes it particularly suited for dealing with model errors in the assimilation cycle, but it also requires an explicit (or operator form) representation of the model error covariance matrix for its implementation. The forcing formulation of WC-4DVar currently implemented at ECMWF augments the 4D-Var control variable with a model error tendency term which is evaluated together with the state control vector during the 4D-Var minimisation and is then used in the subsequent first-guess integration to de-bias the model trajectory:

$$J_{WC}(\boldsymbol{x}_0, \boldsymbol{\eta}) = \frac{1}{2}(\boldsymbol{x}_0 - \boldsymbol{x}_0^b)^T \mathbf{B}^{-1}(\boldsymbol{x}_0 - \boldsymbol{x}_0^b) + \frac{1}{2}\sum_{k=0}^{N}(H(\boldsymbol{x}_k) - \boldsymbol{y}_k)^T \mathbf{R}_k^{-1}(H(\boldsymbol{x}_k) - \boldsymbol{y}_k) +$$

$$\frac{1}{2}(\boldsymbol{\eta} - \boldsymbol{\eta}^b)^T \mathbf{Q}^{-1}(\boldsymbol{\eta} - \boldsymbol{\eta}^b) \qquad (2)$$

$$\boldsymbol{x}_k = M_{k,k-1}(\boldsymbol{x}_{k-1}) + \boldsymbol{\eta}, \quad k = 1,..,N \qquad (3)$$

where $\boldsymbol{\eta}^b$ is the prior estimate of the model error forcing (which are the model error tendencies estimated in the previous WC-4DVar analysis update) and $\mathbf{Q}$ is the model error covariance matrix. While this WC-4DVar formulation has been introduced in operations at ECMWF in 2009, it is only recently (IFS Cycle 47R1, July 2020) that it has been shown to be effective at correcting model systematic errors in the stratosphere (Laloyaux et al., 2020a). The main change in this updated version of the ECMWF WC-4DVar has been the use of a newly estimated model error covariance matrix. This error covariance matrix differs from the previous $\mathbf{Q}$ mainly in terms of horizontal spatial correlation structures, which were tuned to be of significantly larger scale than the corresponding background error covariances ($\mathbf{B}$). This spatial scale separation is necessary for WC-4DVar to work, as the structure of the cost function (2) indicates that if $\mathbf{B}$ and $\mathbf{Q}$ span a similar space model error estimates $\boldsymbol{\eta}$ will alias into initial state analysis increments $\boldsymbol{x}_0$ and vice versa (see Laloyaux et al., 2020b, for an extended discussion of this point; see also Trémolet, 2007, for an illustration of how these problems affected early implementations of WC-4DVar at ECMWF).

While the requirement for scale separation has informed the tuning process of the current operational $\mathbf{Q}$ matrix at ECMWF, the question remains on how to estimate a suitable $\mathbf{Q}$ for use in WC-4DVar in a more objective, less ad-hoc manner. This question is also relevant for other components of the Earth system where assimilation methods based on WC-4DVar are being developed for operational use.

The main purpose of this short note is to describe a possible solution to the problem of $\mathbf{Q}$ estimation based on the use of Artificial Neural Networks (ANN). In the context of model error estimation, Machine Learning (ML) techniques have been the subject of renewed interest in recent years (e.g., Bonavita and Laloyaux, 2020, and references therein). In particular, ANN have been gainfully used to obtain estimates of model error for use in the assimilation cycle and/or during the forecast step. The idea explored in this paper is to extend the use of ANN to estimate the second order moments of the model error and produce a suitable model error covariance matrix for use in WC-4DVar.

This paper is organised as follows. In Sec. 2 we describe and compare the proposed methodology used to estimate the model error covariance with the one currently used in operational practice at ECMWF. In Sec 3 we examine the structure of the model error covariance estimates produced by the two methodologies. In Sec. 4 we examine the impact of using the new ML-derived model error covariance in cycled 4D-Var experiments. In Sec. 5 we discuss these results further in terms of their implications for future research and, more generally, in the context of the current research effort to integrate ML tools in the ECMWF NWP chain. Conclusions are offered in Sec. 6.

## 2) Estimation of Model Error Covariances

*a) Operational Model Error Covariance*

The current approach for the estimation of the model error covariance matrix used in the IFS (which we denote $\mathbf{Q}_{oper}$ in the following) is described in detail in Sec. 3 and 4 of Laloyaux et al., 2020a and its main steps are recalled here for convenience.

The procedure starts with running a predictability experiment in which the ECMWF ensemble forecast system (ENS) is run in a configuration where the initial perturbations are set to zero (i.e., all ensemble members start from the same initial conditions). The forecasts diverge in time due to the presence of stochastic model error perturbations (Stochastic Perturbed Parametrisation Tendency, SPPT, and stochastic kinetic energy backscatter scheme, SKEB; see Leutbecher et al., 2017 for details). The differences from the 12h ensemble forecasts can be used to estimate a model error covariance matrix. This type of **Q** matrix was used in the initial implementation of WC-4DVar (Trémolet, 2007). As noted in Laloyaux et al., 2020, this **Q** matrix (**Q**$_{pred}$ in the following) shows correlation length scales of ~200-250 km in the stratosphere, which are comparable to those of the **B** matrix used in the IFS (Bonavita et al., 2012). This is not surprising, as the ECMWF **B** matrix is itself computed from EDA short range forecast perturbations from the IFS model perturbed with the same type of stochastic model error parameterisations (Bonavita et al., 2012). In the revised **Q** developed in Laloyaux et al., 2020a, (**Q**$_{oper}$), one starts by collecting a number of samples of the estimated model error tendencies (**η**) computed by a WC-4DVar experiment which uses the **Q**$_{pred}$ estimate described above. Under an assumption of model error stationarity, a covariance matrix is computed from these samples. This raw **Q** estimate is then transformed into the final **Q**$_{oper}$ by localising horizontally and vertically and halving the magnitude of its standard deviation profile.

The procedure described above can be seen as a bootstrap method for the estimation of **Q**. Starting with model error perturbations of ~200 km length scale, WC-4DVar acts as a filter and produces model error estimates which lead to a **Q** matrix with significantly larger length scales (~800 km; Laloyaux et al., 2020a, sec. 4). On the other hand, the whole procedure starts from the questionable assumption that the stochastic model error parameterisations used un the EDA and the ENS are appropriate models for the type of slowly evolving, large scale systematic errors that WC-4Dvar is designed to correct for. As shown in Sec. 3, some of the characteristics of the **Q**$_{oper}$ clearly reflect properties of the original stochastic model error parameterisations, which is undesirable. Additionally, successive iterations of the process (i.e., running WC-4DVar with the new **Q** and recomputing the **Q** matrix from the new **η** samples) led to worse results.

*b) ANN-derived Model Error Covariance*

In Bonavita and Laloyaux, 2020, it was shown how to set up and train an Artificial Neural Network that was able to emulate many of the characteristics of the model error estimates produced by WC-4DVar. This ANN implements a nonlinear regression model (see Fig. 1) which is trained to learn cumulated model errors over a 12-hour assimilation window using ECMWF operational analysis increments (A-B) as predictands and a combination of climatological (lat, lon, time_of_day, month) and state-dependent (columns of first-guess forecast fields) predictors. The predicted cumulated errors are then scaled by the length of the assimilation window (12 h) to provide model error tendencies which can be used in WC-4DVar in lieu of (or as a first guess for) the model error tendencies computed during the minimisation (more details in Bonavita and Laloyaux, 2020).

While model error estimates from the ANN model shared many features with those of WC=4DVar, some interesting differences were also apparent. Particularly noticeable was the different structure of model error estimates above ~10 hPa, in the upper stratosphere and mesosphere (Fig. 9 of Bonavita and Laloyaux, 2020a). This has led to the idea of using the trained ANN as a generative model of model error, i.e., use the ANN to generate a representative sample of model errors and compute a new Q matrix ($Q_{ANN}$ in the following) from this sample. In practice, the ANN was run over a 4-month period. The Q matrix computed from this samples has been localised in the horizontal (with a cosine function tapering the correlations to zero between 4000 and 6000 km) in order to remove spurious hemispheric-wide correlations; and in the vertical, with a quadratic function of the distance from the diagonal in order to control sampling noise. The main characteristics of $Q_{ANN}$ are described (and compared with those of the current $Q_{OPER}$) in the following Section 3.

```
Model: "sequential"
_________________________________________________________________
Layer (type)                 Output Shape              Param #
=================================================================
dense (Dense)                (None, 138)               19872
_________________________________________________________________
dropout (Dropout)            (None, 138)               0
_________________________________________________________________
dense_1 (Dense)              (None, 138)               19182
_________________________________________________________________
dropout_1 (Dropout)          (None, 138)               0
_________________________________________________________________
dense_2 (Dense)              (None, 138)               19182
_________________________________________________________________
dropout_2 (Dropout)          (None, 138)               0
_________________________________________________________________
dense_3 (Dense)              (None, 138)               19182
=================================================================
Total params: 77,418
Trainable params: 77,418
Non-trainable params: 0
_________________________________________________________________
```

Figure 1: *Tensorflow model structure for the T and LNSP model error ANN. The ANN is composed of three dense layers with nonlinear (RELU) outputs and a linear output dense layer. More details in Bonavita and Laloyaux, 2020.*

## 3) Diagnosis of Model Error Covariances

In this Section we apply a set of standard diagnostics in order to understand the differences between the current operational Qoper and the ANN-derived Q_ANN and their origin.

*a) Model error profiles*

Some of the more significant differences between $Q_{OPER}$ and $Q_{ANN}$ are in the structure and magnitude of the respective model error tendency profiles (Fig. 2).

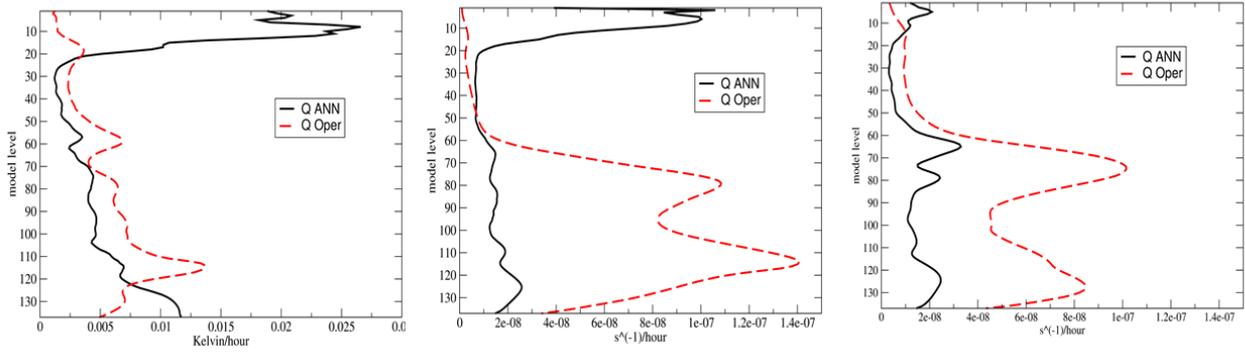

Figure 2: *Vertical profiles of the standard deviation of model error tendencies for the ANN-derived Q (Q$_{ANN}$, black continuous lines) and the operational Q (Q$_{OPER}$, red dashed lines) for temperature (left panel) vorticity (central panel), divergence (right panel). The Q$_{OPER}$ profiles have been rescaled by 0.5 to show the profile actually used in the current ECMWF WC-4DVar. For ease of interpretation, the approx. correspondence between IFS model levels and pressure levels is ml 1 -0.01 hPa; ml 15 – 1 hPa; ml 20 -3hPa; ml 30 -10 hPa; ml 60 -100 hPa; ml 74 -200 hPa; ml 115 -850 hPa; ml 137 - Surface.*

Starting with the temperature model error profiles (Figure 2, left panel), the current operational Q presents decreasing values of error standard deviations in the top 20 model level (0.01 – 3 hPa) and in the lower boundary layer (~925 hPa – Surface). This effect comes from the vertical structure of the stochastic error parameterisations, which are tapered to small values both towards the top and the bottom of the model column, mainly due to stability concerns (Leutbecher et al., 2017). On the other hand, the Q$_{ANN}$ presents relative maxima of model error standard deviations in both the top and bottom model layers, reflecting the presence of known systematic model deficiencies in the sponge layer and in the surface forcing fields (Polichtchouk et al., 2021; Sandu et al., 2021). The operational Q also shows a clear peak in the 850 – 900 hPa layer, which is likely related to the uncertainties generated by the stochastic model error parameterisations on the forecast location of the top of the boundary layer and the related, sharp vertical temperature gradients. This peak is barely visible in the Q$_{ANN}$ temperature error profile.

The model error structures diagnosed by the two Q matrices are also significantly different for the wind variables (vorticity and divergence in the IFS). In particular, the operational Q shows much larger magnitudes (3 to 5 times) than the ANN-derived Q throughout the troposphere and lower stratosphere, which is the atmospheric layer where the stochastic model error parameterisations are more active (Leutbecher et al., 2017). Note however, that these large tropospheric model error values do not affect the current ECMWF WC-4DVar, which is only activated in the stratosphere (Laloyaux et al., 2020a). Conversely, the magnitudes of the diagnosed errors are broadly comparable between $Q_{OPER}$ and $Q_{ANN}$ in the stratosphere and mesosphere, with the notable exception of the top 20 model level for the $Q_{ANN}$ vorticity errors, which, similarly to the temperature model errors, show a remarkable peak 3 to 5 times larger than the values prevalent below. Again, this peak appears connected with systematic deficiencies in the sponge layer of the current IFS model (Polichtchouk et al., 2021).

b)  *Model error vertical correlations*

The structure of vertical correlations of model error is a crucial aspect of the Q matrix, as it determines how model error information is spread in the atmospheric column. As shown in Figure 3, $Q_{OPER}$ and $Q_{ANN}$ show substantial differences. The current model error vertical correlations are nearly homogeneous, have small vertical scales (a vertical localization has been introduced to taper them to zero between 5 and 10 model level distance) and remain positive in that range. The ANN-derived correlations show a more complex structure, particularly for temperature and divergence in the stratosphere/mesosphere. For these two model variables there are also significant negative vertical correlations. Some of these can be physically interpreted (e.g., there is a known change of sign in the IFS model temperature bias in the 60 ~100 hPa layer; also, the convergence-divergence pattern visible in the tropospheric levels of the divergence model errors), which increases confidence in their reliability. Others, e.g., the complex patterns visible in the top 30 model levels for temperature, are more difficult to interpret and could be the result of residual sampling errors.

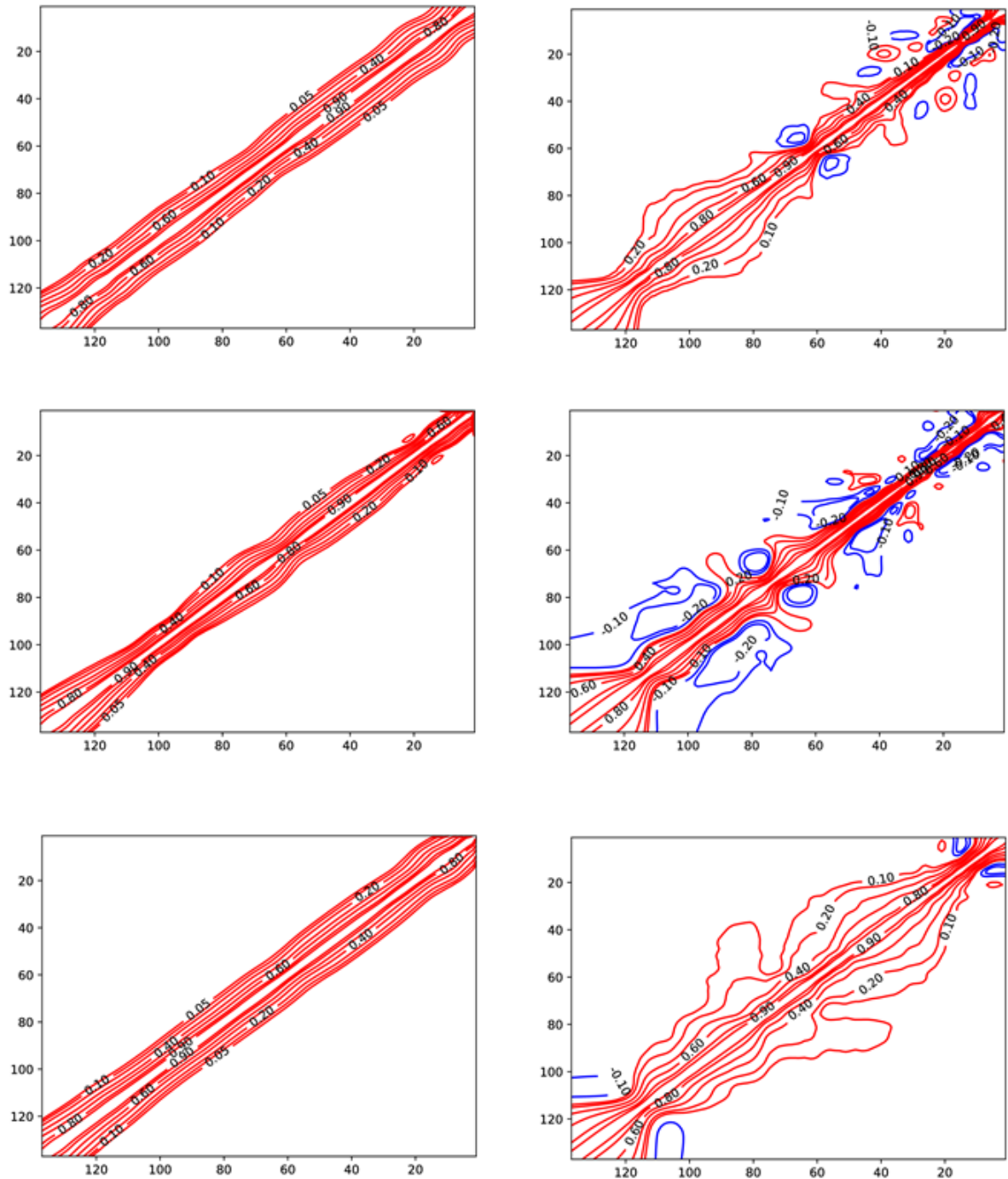

Figure 3: *Vertical correlations of model error tendencies from the operational Q matrix (left column) and the ANN-derived Q matrix (right column). Top row: temperature correlations; middle row: divergence; bottom row: vorticity. Model levels as described in caption of Figure 2.*

c) *Model error horizontal correlations*

The implied model error horizontal correlation structure functions of the new $\mathbf{Q}_{ANN}$ are typically broader by a factor of ~2 than those of the current $\mathbf{Q}_{OPER}$ (Figure 4 for temperature error correlations).

This is partly to be expected, as the ANN from which the new Q has been derived was itself trained on a dataset of low resolutions analysis increments (increments were retrieved at spectral truncation T21, which implies ~900 km grid spacing). The Neural Network has effectively introduced an additional filtering of the error length scales, because a Q sampled directly from a database of T21 analysis increments shows correlations with length scales close to the nominal truncation (not shown). An additional notable aspect is that the change of length scale is largest for the model levels close to the surface, which suggests that the new Q will induce a very different behaviour in WC-4DVar in the boundary layer.

For the wind variables in the model error control vector (vorticity and divergence) the situation is similar, with consistently larger length scales (typically around 700-900 km) than current operational Q (not shown). Largest differences are seen in the boundary layer and close to the tropopause, where current Q has noticeably shorter scales (200-300 km), closer in fact to those of the background error covariance matrix B.

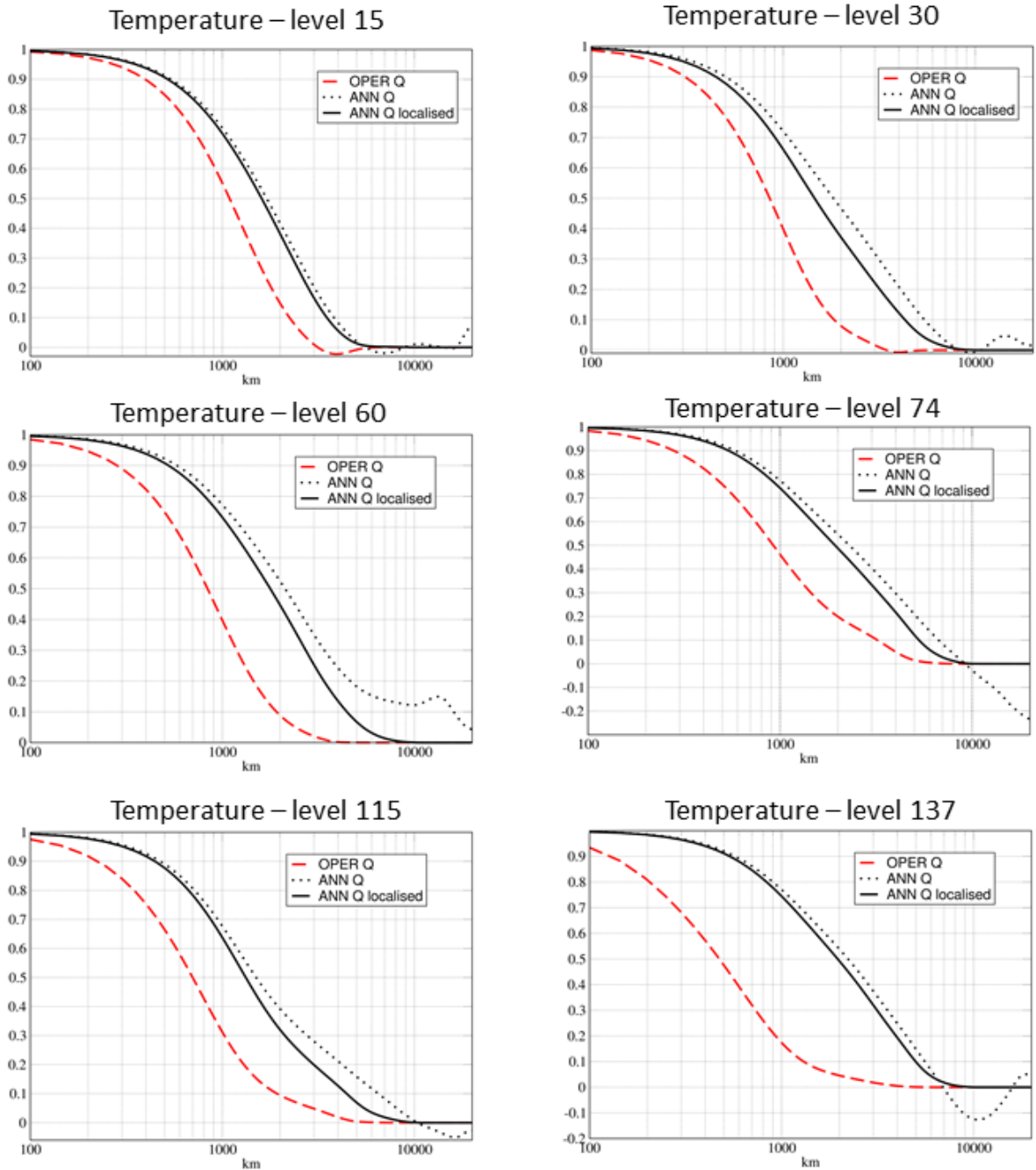

Figure 4: *Model error correlations for temperature implied by the current operational Q (red dashed lines) and ANN-derived Q (black dotted lines for raw Q; black continuous lines for localised Q) at different vertical levels. The approximate correspondence between model levels and pressure levels as in caption of Figure 2.*

**4) Impact on 4D-Var assimilation**

To evaluate the potential impact of the ANN-derived Q we have run a set of WC-4DVar assimilation and forecast experiments conducted with a recent IFS cycle (Cycle 47R2) at resolution TCo399

(approximately 25 km grid spacing) over a period of three months (December 2020 to February 2021). In the following we compare results and diagnostics from a control experiment using the current $Q_{OPER}$ and an experiment using the ANN-derived Q. When useful, we also show results from a strong constraint 4D-Var experiment run over the same period (i.e., where model error estimation and correction is turned off). The WC-4DVar experiments are run in the standard IFS configuration where the model error correction is active above ~100 hPa (model level 60). The extension of WC-4DVar to the full model column is an ongoing research effort at ECMWF.

*a) Model error estimates*

The spatial structure of the model error tendencies estimated during the minimisation process of weak constraint 4D-Var (Eq. 2) depends in large measure on the structure of the assumed model error covariance matrix Q. This is confirmed by the plots in Figure 5, where we present a longitudinal and time average of the temperature model error estimates for the current $Q_{OPER}$ (left panel) and the new $Q_{ANN}$ (right panel). The model error estimates obtained with $Q_{ANN}$ are broadly similar below model level 20 but are significantly larger in the top 20 model levels, reflecting the corresponding spike in the $Q_{ANN}$ temperature standard deviation profile (Figure 2, left panel). Similarly, the model error estimates from the $Q_{ANN}$ WC-4DVar have sharper vertical gradients and are more homogeneous in the horizontal, as the spatial structure of $Q_{ANN}$ vertical and horizontal correlation implies. Similar results are also found for the wind model error estimates (not shown). On the other hand, the standard deviation of the model error estimates over one month is around 10% of their magnitude for both $Q_{OPER}$ and $Q_{ANN}$ (not shown), confirming that both WC-4DVar setups produce error estimates which vary slowly on the timescale of the assimilation window (12 hours).

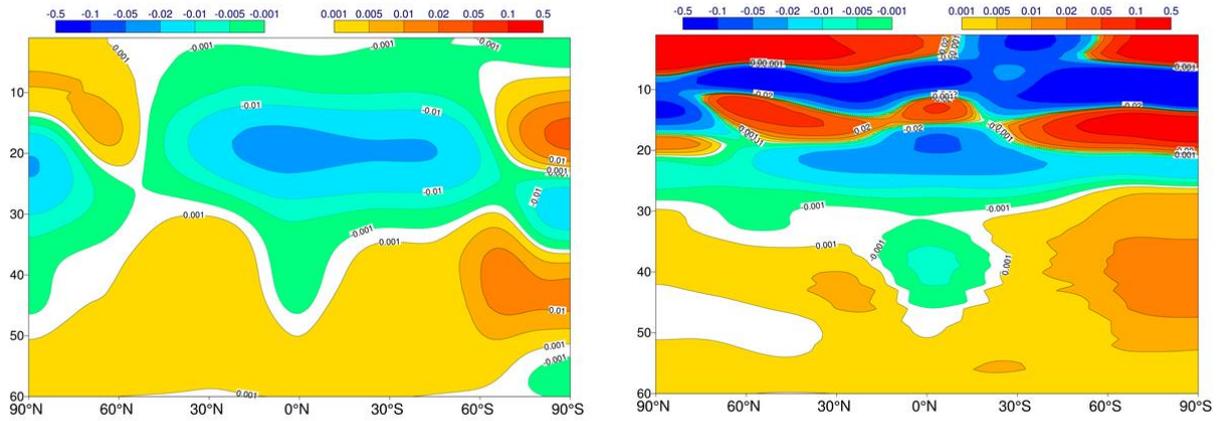

Figure 5: *Longitudinal profile of the temperature model error tendencies estimated by WC-4DVar using the operational Q matrix (left panel) and the ANN-derived Q matrix (right panel). Values averaged over one month (January 2019). Units: Kelvin/hour. Model levels as in the caption of Figure 2.*

b)  *Analysis increments*

The presence of systematic, correlated in time model errors results in the occurrence of systematic analysis increments (Bonavita 2021) and thus one measure of the effectiveness of weak constraint 4D-Var (or any other bias-aware assimilation algorithm) is whether it can reduce or even eliminate systematic analysis increments.

The plots in in Figure 6 confirm that the new ANN-derived Q is effective in reducing the systematic component of analysis increments over the period in question (January 2021). This is directly visible in the mean analysis increment profile for temperature (top row, left panel) where mean increments are a large component of the total analysis increment variance and tend to be homogeneously distributed in the horizontal. For temperature, mean analysis increments are effectively eliminated throughout most of the atmospheric column and reduced to a large extent also in the top ~20 model levels. For wind variables like vorticity (bottom row), global mean increments are very small to start with (note the different scales in the x-axes of the Figure 6 bottom row) and their distribution is not homogeneous. As a consequence, the use of the new $Q_{ANN}$ does not produce a significant effect in terms of globally mean values. However, an impact of the new Q is visible in the reduced magnitude of analysis increments in the higher model levels (Figure 6, bottom row, right panel).

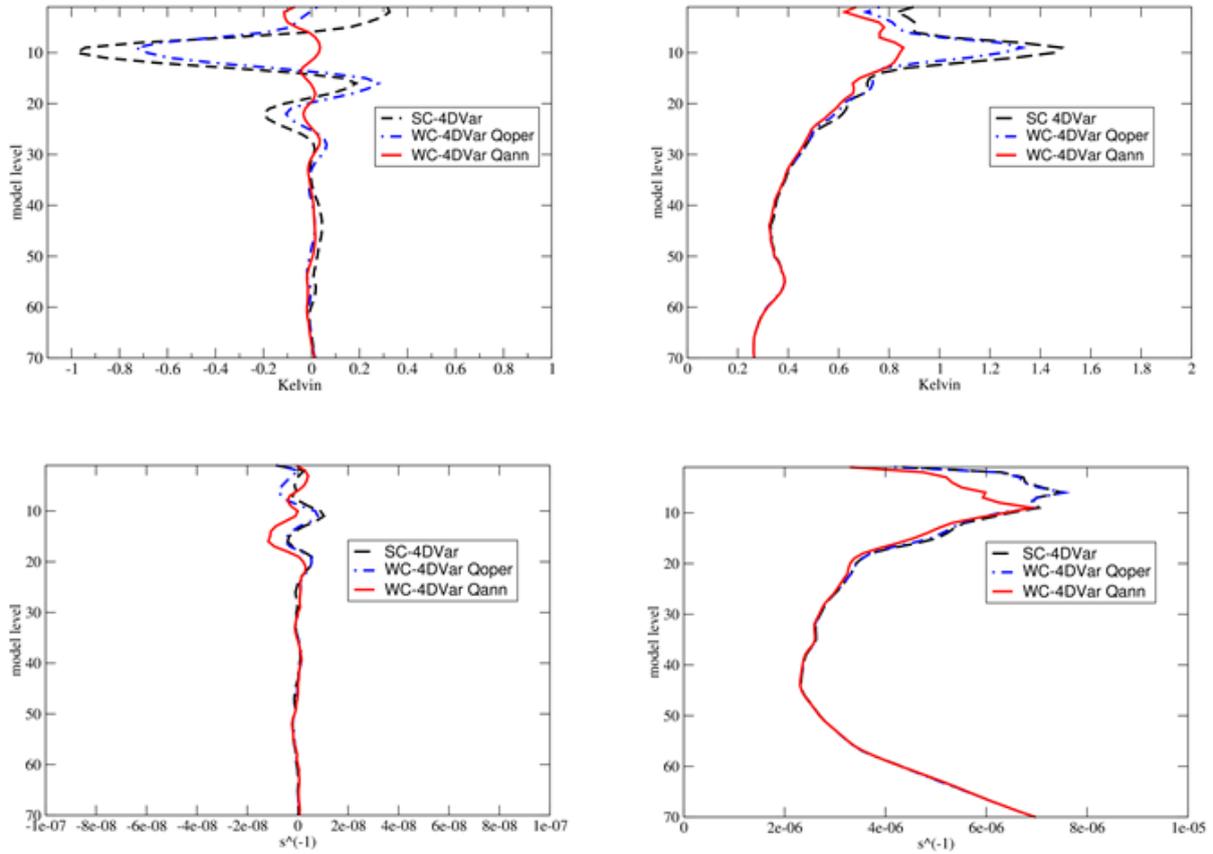

Figure 6: *Vertical profiles of the mean (left column) and rms (right column) analysis increments for temperature (top row) and vorticity (bottom row) averaged over one month (January 2021). Black dashed line: strong constraint 4D-Var; Blue dash-dot line: weak constraint 4D-Var with operational Q; Red continuous line: weak constraint 4D-Var with ANN-derived Q. Model levels as in the caption of Figure 2.*

c)   Analysis and background departures

Analysis and background departures statistics from a cycling data assimilation experiment are one of the main tools to verify the effectiveness of any upgrade to the data assimilation system. Since the new ANN-derived Q produces larger model error estimates in the middle to high stratosphere and mesosphere, we expect analysis and background departures of observations with a sensitivity to these atmospheric layers to be reduced in magnitude. This expectation is generally confirmed by the experimental results. For temperature sounding instruments, a significant reduction in analysis and background observation departures is seen in both microwave and infrared sensors for channels sensitive to stratospheric and mesospheric temperature (Figure 7, top row). The only exception is a slight (~0.2%) degradation for AMSUA channel 9. The weighting function for this channel peaks around 100 hPa, which is difficult layer to handle in the current WC-4DVar setup because this is in the layer where the model error correction is ramped up (model error is linearly activated from ~120 hPa to ~40 hPa); and also, because around 100 hPa the IFS temperature forecast error changes sign.

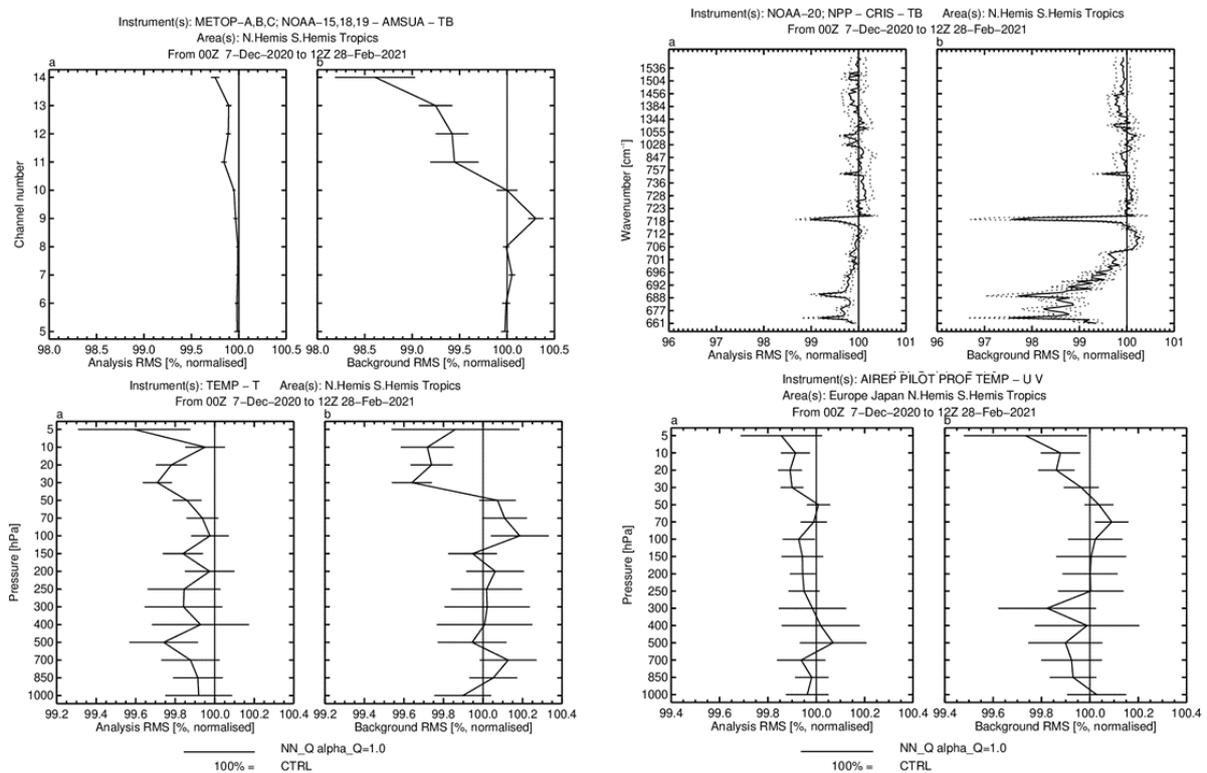

Figure 7: *Relative change in the rms analysis and background fits for: AMSUA microwave sensors (top left panels); CRIS hyperspectral Infrared sensor (top right panels); Radiosonde temperature observations (bottom left panel); Conventional wind observations (bottom right panel). Values lower than 100% indicate that the WC-4DVar experiment using $Q_{ANN}$ has smaller analysis/background departures than the control WC-4DVar using the operational $Q_{OPER}$.*

Similar positive results are also visible for conventional observations (Figure 7, bottom row). Improvements are clearer above 30 hPa, where the new $Q_{ANN}$ is more markedly different from the current $Q_{OPER}$.

Another type of observations that are central in evaluating the performance of the assimilation system are radio occultations (GPS-RO). Their importance stems from the fact that these observations are not bias corrected (contrary to most other satellite-based observations), their nearly homogeneous distribution and their high vertical resolution and reach. It is apparent that the new $Q_{ANN}$ improves root mean squared analysis and background fits to GPS-RO measurements throughout most of the stratosphere, with the exception of the top layer between ~ 45-50 km (~2-1 hPa). In this layer the observation errors assigned to GPS-RO are large (Healy. 2016, Semane et al., 2022), and the analysis is mainly anchored to the highest peaking AMSUA microwave channel 14. As this channel is not bias corrected in the ECMWF 4D-Var, its residual bias affects the model error estimates and the fit to independent observations in the upper layers of the stratosphere.

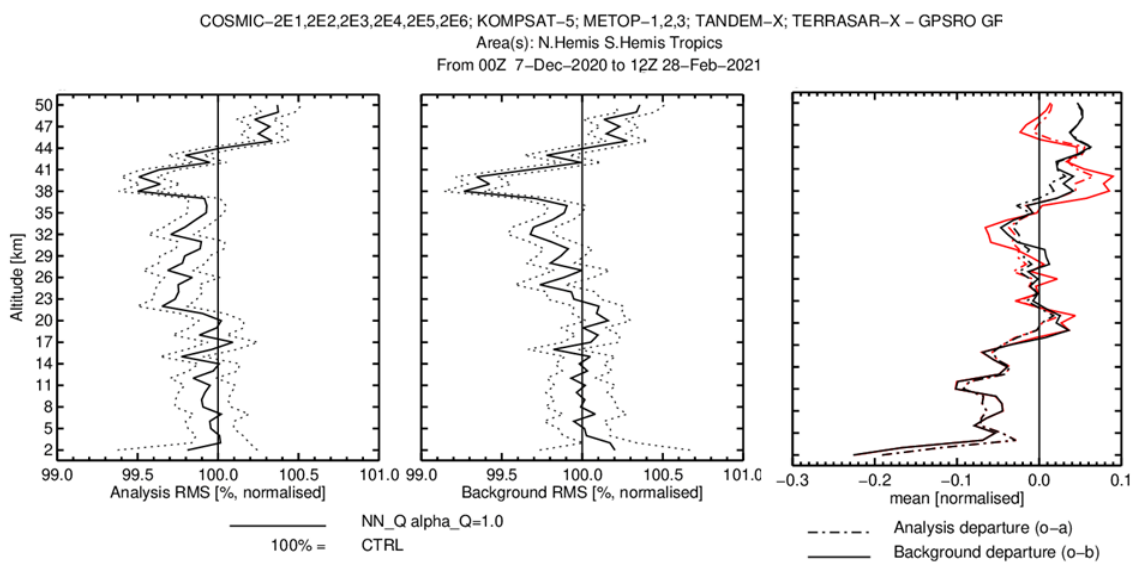

Figure 8 *Left panel: Relative change in the rms analysis fit for GPS-RO observations between WC-4DVar experiment using $Q_{ANN}$ and control experiment using $Q_{OPER}$. Middle panel: as in the left panel for rms background fit. Right panel: vertical profile of mean analysis (dashed lines) and background (continuous lines) GPS-RO departures for WC-4DVar experiment using $Q_{ANN}$ (black lines) and control experiment using $Q_{OPER}$. (red lines).*

d) *Forecast impact*

In the WC-4DVar configuration used in the experiments described here model error is only active in the stratosphere above ~ 100 hPa. In this atmospheric layer the new $Q_{ANN}$ produces model error estimates that are significantly different only from model level 30 (~ 10 hPa, see Figure 5), so we can expect forecast impact to be broadly confined in the middle to high stratosphere. This is confirmed by forecast skill verification (Figure 9). Elsewhere results are neutral with when verified against radiosondes. While radiosonde verification is the most appropriate for a data assimilation system change designed to improve the treatment of model errors, it is also limited by the fact that radiosonde measurements are generally not available above 10 hPa. Verification against the operational ECMWF analyses shows similar results as the radiosonde verification up to 10 hPa. Above 10 hPa, results appear generally neutral, but a degradation is visible in the temperature forecast in the extra-tropics in the 1-3 hPa layer (not shown). As this is the layer where observation departures against GPS-RO measurements were also negatively affected by the use of the new $Q_{ANN}$ (Figure 8), this appears to be a genuine signal.

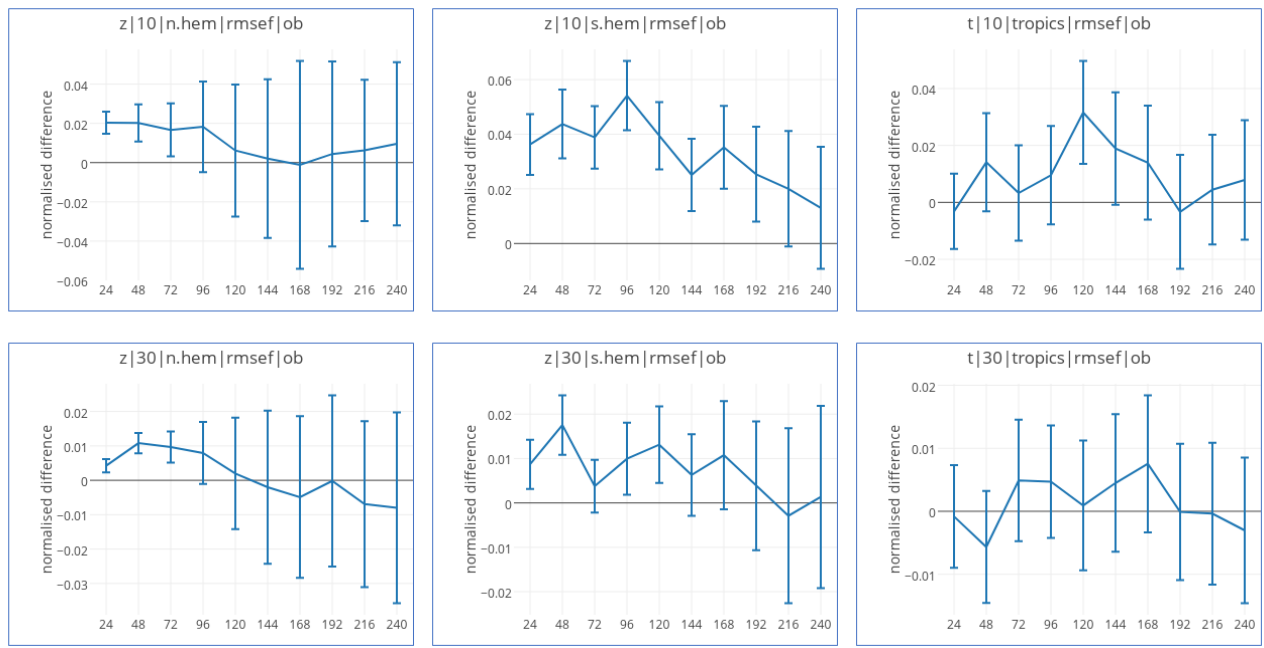

Figure 9 *Relative difference in forecast RMS errors at 10 hPa (top row) and 30 hPa (bottom row) for: a) geopotential height in the northern extra-tropics (left column); b) geopotential height for the southern extra-tropics (middle column; c) temperature in the tropics (right column). Values above the zero line indicate higher skill (lower rms errors) of the $Q_{ANN}$ WC-4DVar with respect to the $Q_{OPER}$ WC-4DVar. Verification with respect to radiosonde observations over a three-months period.*

5) Discussion

The idea we have explored here is to use an ANN as a generative model to sample the model error distribution of the IFS and provide a model error covariance matrix Q suitable for use in a cycling

WC-4Dvar framework. The resulting Q is significantly different from the current Q, which reflects to a large extent features inherited from the stochastic model perturbation schemes used in the ECMWF EDA and ensemble forecasts, and closer to the structure of the mean analysis increments. Note, however, that simply using a climatology of analysis increments to build a Q matrix is not sufficient, as the resulting Q matrix is structurally different from the ANN-derived Q (not shown). In fact, it has been impossible to evaluate its performance due to repeated 4D-Var and forecast model crashes that we encountered when trying to use it in a cycling WC-4DVar experiment.

The results of a limited set of WC-4DVar experiments using $Q_{ANN}$ indicate that it is more effective than $Q_{OPER}$ to reduce systematic analysis increments and, in general, reduce the magnitude of the analysis increments in a rms sense. For temperature in particular, globally averaged mean temperature analysis increments are almost eliminated throughout the stratosphere.

Comparison with observations in the ECMWF data assimilation system generally confirms the improved behaviour of WC-4DVar with QANN. The analysis and first guess fits to stratospheric sounding satellite radiances are generally improved, particularly for higher-peaking channels, together with a reduction in the bias correction coefficients for some of these channels. Conventional observations also show improved observation fits in the stratosphere. Two exceptions have also been identified. One is connected to a very small degradation in satellite and conventional observation fits around the 70-100 hPa layer. This is due to the fact that model error corrections are tapered down in that area where the IFS model temperature error also changes sign. Preliminary sensitivity studies indicate that either the extension of the model error correction to the whole troposphere and/or stricter localisation of the model error covariances in the 70-100 hPa layer are able to get rid of the problem. The other problem identified in the experimentation has been the degradation of GPS-RO analysis and background forecast departures in the 50-45 km layer (~1-3 hPa), mostly due to the increase in the mean departures. This degradation appears to be due to the anchoring effect of the top-peaking microwave observations (AMSUA channel 14, predominantly), which are not bias corrected in the IFS data assimilation system. In the same 45-50 km layer the assumed GPS-RO observation errors are currently inflated to very large values in the IFS assimilation system (and in most other global NWP Centres, e.g., Semane et al., 2022), so they have very limited impact on the analysis and model error estimates. The consequence is that model error estimates are drawn towards further reducing observation departures for AMSUA top channels (e.g., Figure 7) and thus absorb residual observation biases from these observations. Current developments at ECMWF (Noureddine Semane, personal communication) aim to increase the weight given to GPS-RO observations in the analysis cycle and also to increase the vertical range of GPS-RO usage. We expect that these developments will contribute to resolve this issue and also lead to a reconsideration on the choice of anchoring (i.e., not bias-corrected) observations in the stratosphere, and, more generally, to the design of the ECMWF observation variational bias correction (VarBC) system. One aspect that will merit renewed attention is the choice of predictors of the AMSUA observation biases. This currently include an air-mass dependent component (Han and Bormann, 2016) which could potentially hide a significant model error component in atmospheric layers where the analysis is not sufficiently constrained by other observing systems (e.g., middle to high stratosphere).

Building on the success of this initial effort, various further developments can be envisaged. The most obvious is the extension of the WC-4DVar to the whole atmospheric column. This is a challenging problem for several reasons. To start with, model biases in the troposphere are smaller both in an absolute sense and relative to model variability; as a consequence, there is a heightened possibility that residual observation biases are aliased by WC-4DVar into model error estimates. On the upside, getting rid of the transition zone where model error estimates are tapered down in the lower stratosphere would help get rid of spurious effects in this layer (and possibly improve things elsewhere too, due to interactions with observations sensing deep atmospheric layers). The new $Q_{ANN}$ presents remarkably different error standard deviation profiles in the whole boundary layer compared to the current $Q_{OPER}$, and it appears to avoid some of the artefacts introduced in the $Q_{OPER}$ profiles by the stochastic model uncertainty schemes. This is currently being evaluated.

Another important development area is the extension of the model error control vector used in WC-

4DVar, which is currently limited to temperature, lnsp, vorticity and divergence. This will initially focus on humidity and ozone, which are both known to suffer from non-negligible model systematic errors. The use of ANN to build suitable model error covariance matrices for these quantities as described in this paper may provide a principled and relatively easy path to achieve this.

The current model error covariance model is defined in spectral space (i.e., Courtier et al., 1998). This has many advantages (e.g., compact, non-separable) but the disadvantage of global homogeneity, which could be a limiting factor especially in the troposphere. Further along the horizon, we envisage to upgrade to a more sophisticated representation of the model error covariance, based on the wavelet formulation already used for the background error covariance matrix (Bonavita et al., 2016).

# REFERENCES

Boisserie, M., Arbogast, P., Descamps, L., Pannekoucke, O. and Raynaud, L. (2014), Estimating and diagnosing model error variances in the Météo-France global NWP model. Q.J.R. Meteorol. Soc., 140: 846-854. https://doi.org/10.1002/qj.2173

Bonavita, M., Isaksen, L. and Hólm, E., 2012: On the use of EDA background error variances in the ECMWF 4D-Var. Q.J.R. Meteorol. Soc., 138: 1540-1559. https://doi.org/10.1002/qj.1899

Bonavita, M., Hólm, E., Isaksen, L. and Fisher, M. (2016), The evolution of the ECMWF hybrid data assimilation system. Q.J.R. Meteorol. Soc., 142: 287-303. https://doi.org/10.1002/qj.2652

Bonavita, M., & Laloyaux, P., 2020: Machine learning for model error inference and correction. Journal of Advances in Modeling Earth Systems, 12, e2020MS002232. https://doi.org/10.1029/2020MS002232

Bowler, N. E., 2017: On the diagnosis of model error statistics using weak-constraint data assimilation. Quart. J. Roy. Meteor. Soc., 143, 1916–1928, https://doi.org/10.1002/qj.3051.

Courtier, P, Andersson, E, Heckley, WA, Pailleux, J, Vasiljevic, D, Hamrud, M, Hollingsworth, A, Rabier, F, Fisher, M, 1998: The ECMWF implementation of three dimensional variational assimilation. Part 1: Formulation. ECMWF Tech. Mem. n. 241. Doi: 10.21957/unhecz1kq

Daley R. (1992), Estimating model error covariances for application to atmospheric data assimilation. Mon. Weather Rev. 120: 1735–1746.

ECMWF, 2020. Fact sheet: ECMWF's use of satellite observations. Available at https://www.ecmwf.int/en/about/media-centre/focus/2020/fact-sheet-ecmwfs-use-satellite-

observations

Fisher M. 2003. 'Background-error covariance modelling'. In Proceedings of seminar on recent developments in data assimilation for atmosphere and ocean. ECMWF: Reading, UK. 45–63.

Han, W. and N. Bormann, 2016. Constrained adaptive bias correction for satellite radiance assimilation in the ECMWF 4D-Var system. ECMWF Tech. Mem. n. 783. Doi: http://dx.doi.org/10.21957/rex0omex

Healy, S., 2016. Estimates of GNSS radio occultation bending angle and refractivity error statistics. ROM SAF Report 26. Available at https://www.romsaf.org/general-documents/rsr/rsr_26.pdf

Houtekamer P, Lefaivre L, Derome J, Ritchie H, Mitchell HL. 1996. A system simulation approach to ensemble prediction. Mon. Weather Rev. 124:1225–1242.

Laloyaux, P, Bonavita, M, Dahoui, M, et al. Towards an unbiased stratospheric analysis. Q J R Meteorol Soc. 2020a; 146: 2392– 2409. https://doi.org/10.1002/qj.3798

Laloyaux, P., Bonavita, M., Chrust, M., Gürol, S. Exploring the potential and limitations of weak-constraint 4D-Var. Q J R Meteorol Soc. 2020b; 146: 4067– 4082. https://doi.org/10.1002/qj.3891

Leutbecher, M., Lock, S.-J., Ollinaho, P., et al. (2017). Stochastic representations of model uncertainties at ECMWF: State of the art and future vision. Quarterly Journal of the Royal Meteorological Society, 143(707), 2315– 2339.

Polichtchouk, I., P. Bechtold, M. Bonavita, R. Forbes, S. Healy, R. Hogan, P. Laloyaux, M. Rennie, T. Stockdale, N. Wedi, M. Diamantakis, J. Flemming, S. English, L. Isaksen, F. Vána, S. Gisinger, and N. Byrne, 2021: Stratospheric modelling and assimilation. ECMWF Tech. Mem. n. 877, DOI: https://10.21957/25hegfoq

Sandu, I, Haiden, T, Balsamo, G, Schmederer, P, Arduini, G, Day, J, Beljaars, A, Ben-Bouallegue, Z, Boussetta, S, Leutbecher, M, Magnusson, L, de Rosnay, P, 2020: Addressing near-surface forecast biases: outcomes of the ECMWF project 'Understanding uncertainties in surface atmosphere exchange' (USURF). ECMWF Tech. Mem. n. 875, DOI: https://10.21957/wxjwsojvf

Semane, N., Anthes, R., Sjoberg, J., Healy, S., & Ruston, B. (2022). Comparison of Desroziers and Three-Cornered Hat Methods for Estimating COSMIC-2 Bending Angle Uncertainties, Journal of Atmospheric and Oceanic Technology, 39(7), 929-939. Retrieved Sep 22, 2022, from https://journals.ametsoc.org/view/journals/atot/39/7/JTECH-D-21-0175.1.xml

Tandeo, P., Ailliot, P., Bocquet, M., Carrassi, A., Miyoshi, T., Pulido, M., & Zhen, Y. (2020). A Review of Innovation-Based Methods to Jointly Estimate Model and Observation Error Covariance Matrices in Ensemble Data Assimilation, Mon. Weather Rev. 148(10), 3973-3994. Retrieved Aug 19, 2022, from https://journals.ametsoc.org/view/journals/mwre/148/10/mwrD190240.xml

Todling R. 2014. A lag-1 smoother approach to system error estimation: Sequential method. Q. J. R. Meteorol. Soc., doi: 10.1002/qj.2460.

Trémolet Y. 2006. Accounting for an imperfect model in 4D-Var. Q. J. R. Meteorol. Soc. 132: 2483–2504, doi: 10.1256/qj.05.224.

Trémolet Y. 2007. Model-error estimation in 4D-Var. Q. J. R. Meteorol. Soc. **133**: 1267–1280. https://doi.org/10.1002/qj.94.